\pdfoutput=1
\documentclass[conference,a4paper,10pt]{IEEEtran}
\usepackage[latin1]{inputenc}
\usepackage{times,amsmath}
\usepackage{amsfonts}
\usepackage{pstool}
\usepackage{subfigure}
\usepackage{multirow}
\usepackage{multicol}
\usepackage{enumerate}
\usepackage{graphicx}
\usepackage{mathtools, cuted}
\usepackage{MnSymbol}
\usepackage{stfloats}
\usepackage[table]{xcolor}
\usepackage{diagbox}
\usepackage{slashbox}
\usepackage{nohyperref}
\usepackage{algorithm,algorithmic}
\usepackage{bigints}
\usepackage{amsmath}
\usepackage{cite}
\begin{document}
\title{A Deep Learning \& Fast Wavelet Transform-based Hybrid Approach for Denoising of PPG Signals}
\author{
\IEEEauthorblockN{Rabia Ahmed$^\ast$, Ahsan Mehmood$^\ast$,  M.\ Mahboob\ Ur\ Rahman$^\ast$, and  Octavia\ A.\ Dobre$^\dagger$  \\
$^\ast$Electrical engineering department, Information Technology University, Lahore, Pakistan \\
$^\dagger$Department of Electrical \& Computer engineering, Memorial University of Newfoundland, St. Johns, Canada \\
$^\ast$mahboob.rahman@itu.edu.pk, $^\dagger$odobre@mun.ca \\ 
}
}
\maketitle


\begin{abstract}

This letter presents a novel hybrid method that leverages deep learning to exploit the multi-resolution analysis capability of the wavelets, in order to denoise a photoplethysmography (PPG) signal. Under the proposed method, a noisy PPG sequence of length $N$ is first decomposed into $L$ detailed coefficients using the fast wavelet transform (FWT). Then, the clean PPG sequence is reconstructed as follows. A custom feedforward neural network (FFNN) provides the binary weights for each of the wavelet sub-signals outputted by the inverse-FWT block. This way, all those sub-signals which correspond to noise  or artefacts are discarded during reconstruction. The FFNN is trained on the Beth Israel Deaconess Medical Center (BIDMC) dataset under the supervised learning framework, whereby we compute the mean squared-error (MSE) between the denoised sequence and the reference clean PPG signal, and compute the gradient of the MSE for the back-propagation. Numerical results show that the proposed method effectively denoises the corrupted PPG and video-PPG signal. 


\end{abstract}

\section{Introduction}

Electrocardiogram (ECG) remains the golden standard to study the physiology of the heart, to date. Nevertheless, the limitations of ECG, e.g., placement of electrodes on patient's chest and limbs by a trained professional, requirement of no chest movement during the experiment, etc., have inspired the researchers to explore and develop alternate methods as well. To this end, photoplethysmography (PPG) is a low-cost, non-invasive, optical technique that has recently gained widespread acceptance for (do-it-yourself) monitoring of cardiac activity. The off-the-shelf PPG devices (pulse oximeters, wearable smart bands, smart watches) are typically attached to the peripheral sites (e.g., wrist, finger, ear, etc.) to monitor various important physiological parameters, e.g., heart rate, heart rate variability, blood Oxygen saturation level (SpO2), and respiratory rate~\cite{A2}. More recently, the PPG signal has been used to aid in the diagnosis of several cardiac vascular, respiratory, and neurological diseases~\cite{A3}. 

\begin{figure}
    \centering
     \subfigure[]
    {\includegraphics[width=0.24\textwidth]{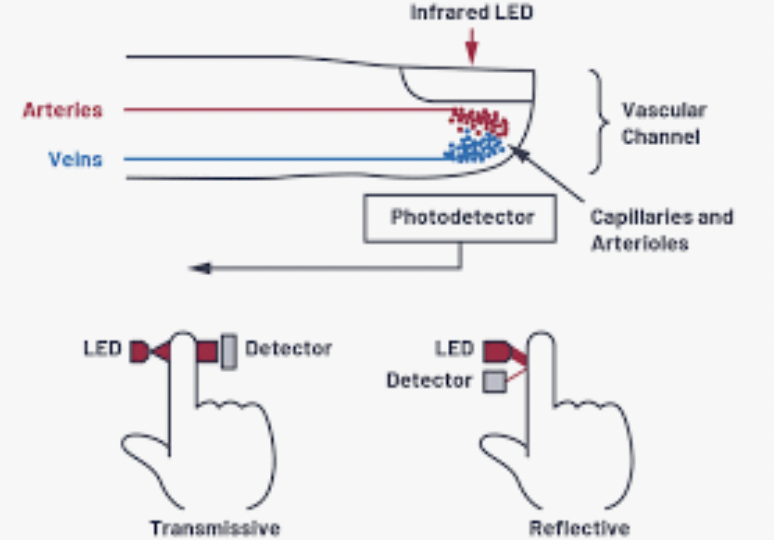}} 
   \subfigure[]
   {\includegraphics[width=0.24\textwidth]{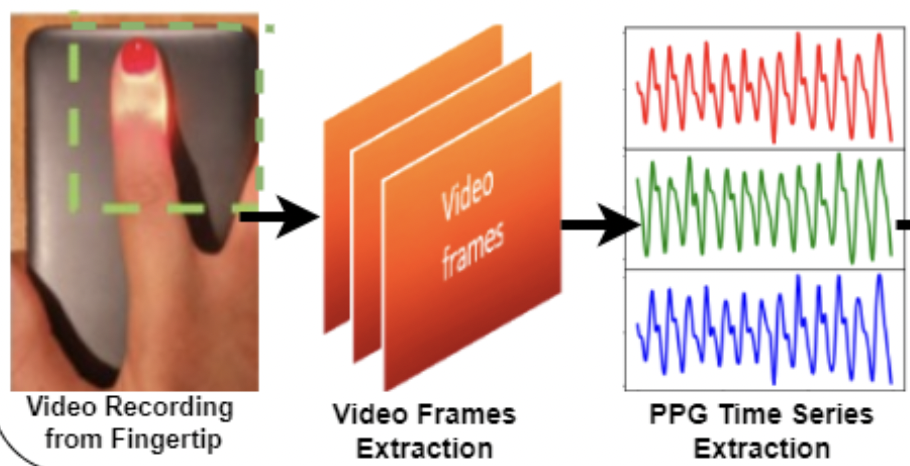}} 
    \caption{(a) Oximeter-based PPG, (b) Smartphone-based video-PPG. }
    \label{fig:1}
\end{figure}

A PPG device utilizes light at two different wavelengths (typically, red and infrared) to measure the volumetric changes in blood flow inside blood vessels at the distal sites. That is, a PPG device is composed of a light emitting diode (LED) that emits (red and infrared) light, and a photodetector (PD) that collects the fraction of light reflected from the skin tissue. Two kinds of PPG devices are prevalent: transmissive and reflective. In transmissive type (oximeters), LED and PD are on the opposite side of the skin tissue, the LED impinges the light onto the skin tissue while the PD collects the light that penetrates through the tissue. In reflective type (smart watches), the PD is placed besides the LED on the same side of the skin tissue (see Fig. \ref{fig:1} a). More recently, video-PPG has started to attract the attention of the research community whereby one places his/her fingertip on the smartphone camera in order to record a small video which is then processed to extract the PPG signal (see Fig. \ref{fig:1} b).  

Having motivated the PPG techniques, there are certain challenges associated with the PPG waveform too, e.g., person-specific penetration depth of the light incident onto the skin tissue~\cite{A5}, variation in accuracy from one body-site to another, interference from nearby power-lines, baseline wander noise ~\cite{A6}, motion artefacts due to ambient light and due to voluntary or involuntary movement of the subject (e.g., during exercise, sleep). 
Therefore, in order to make sure that the accuracy of the body vitals returned by a PPG device is not compromised, there is a pressing need to design efficient denoising algorithms to remove unwanted noise and motion-induced artefacts (e.g., baseline drift and spikes) from a noisy PPG signal. This is precisely the agenda of this work. 

{\bf Contributions.}
The proposed method consists of two distinct phases. During the decomposition phase, the fast wavelet transform (FWT) is utilized to extract multiple approximation and detailed coefficients from a noisy PPG signal. During the reconstruction phase, a custom feedforward neural network (FFNN) systematically selects the (noise-free) wavelet sub-signals outputted by the inverse-FWT in order to help synthesize a clean PPG signal.
To train the FFNN in a supervised manner, gradients of mean squared error (MSE) loss function are computed for the backpropagation purposes.

{\bf Outline.}
The rest of this paper is organized as follows. Section II summarizes the selected related work. Section \ref{NN} provides key details of our proposed DL-assisted hybrid denoising method. Section IV provides selected numerical results. Finally, Section \ref{conc} concludes the paper.

\section{Related Work}

Due to space-constraint, this section summarizes only selected related works, as presented below.\footnote{For instance, there are works that design adaptive filters, perform singular value decomposition, or use the multi-scale principle component analysis for PPG denoising. These are not referenced due to the lack of space. }

The Fourier series analysis and bandpass filtering \cite{A7}, iterative singular spectrum-based method \cite{A9}, least mean square algorithm \cite{A12}, empirical mode decomposition combined with spectrum subtraction \cite{A13}, spectral matrix decomposition \cite{A14}, independent component analysis combined with wavelet \cite{A22}, discrete wavelet transform (DWT) \cite{A15}, multi-level DWT \cite{A16}, DWT along with adaptive filtering \cite{A17}, and periodic averaging filtering \cite{A8} are some of the techniques used for noise and artefact removal from PPG signals. 


Recently, machine and deep learning based works on PPG denoising have emerged. For instance, \cite{A20} utilizes the principal component analysis along with an artificial neural network, while \cite{A1} considers a cycle generative adversarial network for noise removal from PPG signals.



\section{The Proposed Method}\label{NN}




This section first introduces the PPG denoising problem as an optimization program. Then, the key details of the proposed deep learning (DL) and wavelets-based hybrid method are outlined, as well as the architecture of the custom DL model. 

\subsection{Problem Formulation}

\begin{figure} 
\centering
\includegraphics[width=3.5in]{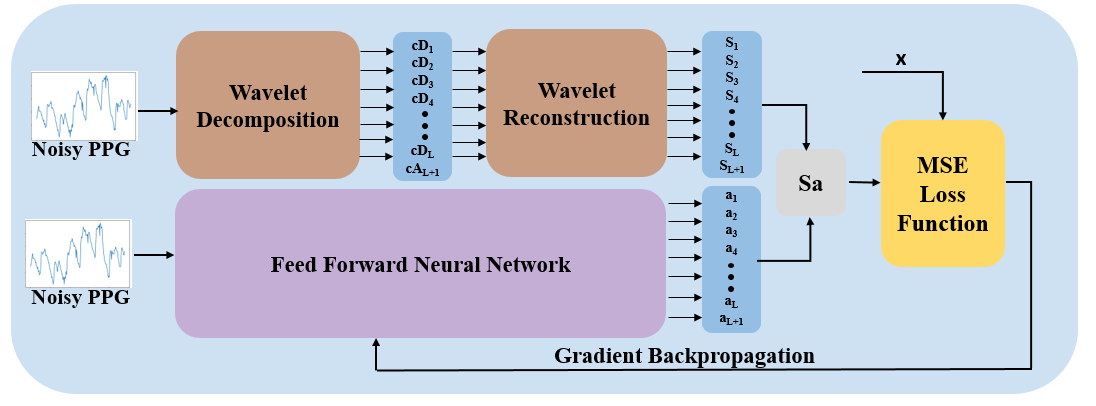}
\caption{The proposed deep learning \& the wavelets-based hybrid method. }
\centering
\label{fig:SM}
\end{figure}

The raw PPG signal acquired by a PPG device is typically contaminated by: the respiratory-induced baseline, hardware-induced high-frequency noise, artefacts due to motion and ambient light. Thus, in order to denoise the raw noisy PPG signal, we begin by taking the FWT of the noisy PPG signal in order to decompose the signal into ($L$) detail coefficients and one approximate coefficient (see Fig. \ref{fig:SM}). 


\begin{figure} 
\centering
\includegraphics[width=2.7in]{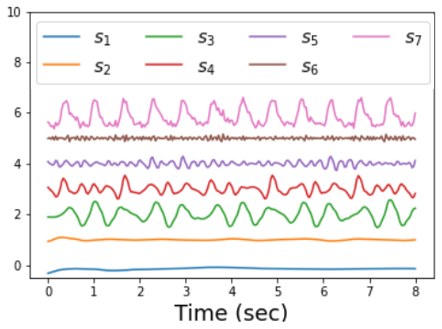}
\caption{Reconstruction of sub-signals by the wavelet reconstruction block.}
\centering
\label{fig:sub_sig}
\end{figure}

The output of the wavelet reconstruction block consists of the sub-signals $\mathbf{s}_i$ that together satisfy the following equation:
\begin{equation}
\label{eq:2}
    \textbf{y} =  \textbf{s}_{1}  + \textbf{s}_{2} +... + \textbf{s}_{L+1},
\end{equation}
where  $\textbf{s}_i \in \mathcal{R}^{N \times 1}$  is the sub-signal reconstructed from the $i$-th detailed coefficient (while setting all other coefficients to 0). $\textbf{s}_{L+1}$ is the sub-signal obtained using the approximate coefficient only (with all detailed coefficients being 0). For a quick illustration, Fig. \ref{fig:sub_sig} shows a few such sub-signals ($\mathbf{s}_1$ to $\mathbf{s}_7$) outputted by the wavelet reconstruction block. 

Equation (\ref{eq:2}) suggests that the noisy PPG signal $\textbf{y}$ can be denoised by selectively suppressing some sub-signals. This implies that the denoised PPG signal can be written as follows:
\begin{equation}
\label{eq:3}
    \hat{\textbf{x}} = a_{1} \textbf{s}_{1}  +a_{2} \textbf{s}_{2} +... +a_{L+1} \textbf{s}_{L+1},
\end{equation}
where $a_i$ equals $0$ for the sub-signals $\textbf{s}_i$ that correspond to the noise and artefacts, and $1$ for the sub-signals which (when summed together) yield the clean PPG signal. Thus, the denoising problem boils down to the optimal selection of $a_i$'s. Let us denote $\textbf{a} = [a_1, a_2, ..., a_{L+1}]^T$
and $\mathbf{S} = [\textbf{s}_1, \textbf{s}_2, ... , \textbf{s}_{L+1}]$,
then we can rewrite (\ref{eq:3}) as follows:
\begin{equation}
\label{eq:4}
    \hat{\textbf{x}} = \mathbf{S}\textbf{a}.
\end{equation}
To obtain the denoised signal $\hat{\textbf{x}}$, the optimal $\textbf{a}$ can be obtained by solving the following optimization problem:

\begin{equation}
\begin{aligned}
\min_{a_{1},a_{2} ... a_{L+1} } \quad & || \mathbf{S} \mathbf{a} -\mathbf{x} ||_2^2 \\
\textrm{s.t.} \quad & {a_{i}} \in \{0,1\}, i = 1,2,...,L+1 \\
\end{aligned}
\end{equation}
where $\mathbf{x}$ is the clean PPG signal. The binary integer constraint makes this optimization problem an integer linear program that is difficult to solve. Thus, we relax the integer constraint by allowing $a_i$ to take any value between $0$ and $1$. It helps since the constraint can now be satisfied by our DL model by using the sigmoid activation function at the output layer. Thus, we reach the following optimization problem (which also serves as the loss function for our DL model):


\begin{equation}
\begin{aligned}
\min_{a_{1},a_{2} ... a_{L+1} } \quad & || \mathbf{S} \mathbf{a} -\mathbf{x} ||_2^2 \\
\textrm{s.t.} \quad & 0 \leq {a_{i}} \leq 1, i = 1,2,...,L+1 \\
\end{aligned}
\end{equation}

\subsection{Proposed DL and Wavelet-based Hybrid Approach}
Under the proposed method, the noisy PPG signal goes as input to the wavelet decomposition block, while the wavelet reconstruction block returns the sub-signals $\mathbf{s}_i$ that satisfy (\ref{eq:2}) (see Fig. \ref{fig:SM}). The noisy PPG signal is also fed to the DL model in parallel, which generates the optimal weight vector $\textbf{a}$. The product of the matrix $\mathbf{S}$ (generated by the wavelet reconstruction block) with the vector $\textbf{a}$ (generated by the DL model) is the denoised signal. The MSE between the denoised signal and the clean PPG signal is utilized as a loss function to compute the gradients, which are then backpropagated in order to train the FFNN in a supervised manner.

\begin{figure} 
\centering
\includegraphics[width=2.9in]{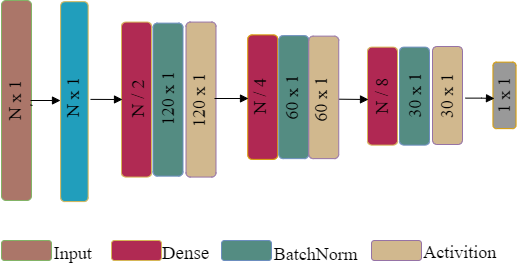}
\caption{Architecture of the custom FFNN.}
\centering
\label{fig:NN}
\end{figure}

\textit{The FFNN Architecture:}
The DL model used in the proposed method is an FFNN (see Fig. \ref{fig:NN}). It consists of an input layer of size $N$, three hidden layers, and an output layer. The number of neurons successively halve from one hidden layer to the next, with $L+1$ neurons in the output layer. Each hidden layer consists of a dense layer followed by batch normalization and activation (which in our case is ReLu). Finally, the output layer utilizes the sigmoid activation function.  

\subsection{Performance Metrics}
The performance of the proposed method is evaluated using the following two performance metrics: 
MSE: $= \frac{1}{M} \sum_{j=1}^{\ M} (\textbf{x}_j-\hat{\textbf{x}}_j)^2$, and peak signal-to-noise ratio (PSNR): $= 10  {\log{_{10}}} (1/\textrm{MSE})$ in decibels, where $\textbf{x}_j$ is the clean PPG signal, $\hat{\textbf{x}}_j$ is denoised PPG signal, and $M$ is the batch size. 



\section{Performance Evaluation}
\label{performance}

We first provide key details about the training of the custom FFNN presented in Fig. 4, followed by a brief description of the two simulation experiments, and finally the results.

\textit{Training of the custom FFNN.}
All the trainable parameters are initialized by Xavier initializer. The model is optimized using Adam optimizer with an initial learning rate (LR) of 0.001. To avoid the over-fitting, an early stopping is applied. The details of other hyper-parameters are given in Table \ref{tab:table3}.

\begin{table}
\caption{\label{tab:table3}Hyper-parameters for the custom FFNN.}
\begin{center}
\begin{tabular}{c*{2}{c}}
\\\hline
Hyper-parameter & Value \\\hline\hline
Training dataset size & $3873$\\\hline
Test dataset size & $240$ \\\hline
Training batch size & $100$ \\\hline
Validation batch size & $100$ \\\hline
Number of epochs & $500$ \\\hline
\end{tabular}
\end{center}
 \end{table}
    
We now describe each of the two experiments performed to validate the performance of the proposed method.

\begin{figure} 
\centering
\includegraphics[width=3.2in]{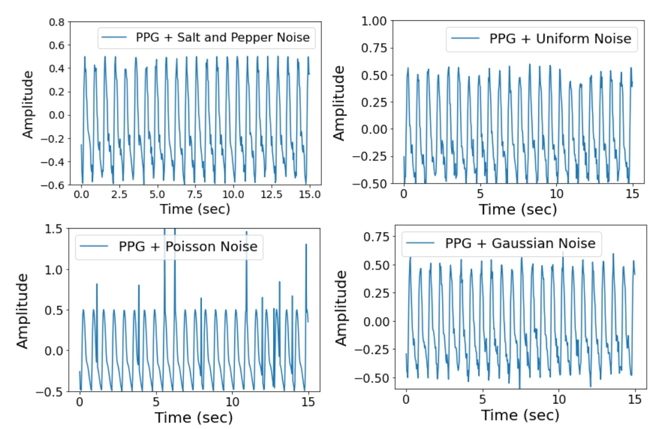}
\caption{Some noisy PPG signals.}
\centering
\label{fig:Noisy PPG}
\end{figure}

\textit{Experiment I.}
The first experiment is a controlled one. We fetch clean PPG signals from the BIDMC dataset \cite{BIDMC}, and corrupt them by adding four types of noises which are known to degrade the PPG signals quite often. Specifically, we consider the following noise types: Gaussian (zero-mean, variance$=0.05$), Poisson (with $\lambda =0.02$), uniform ($[ -0.1,0.1]$), and salt-and-pepper noise (with a density$=0.05$). Fig. \ref{fig:Noisy PPG} shows a snapshot of the resulting noisy PPG signals.

Next, keeping in mind the multi-level decomposition capability of the FWT, we investigate the impact of the depth of decomposition on the denoising performance of the proposed method. Fig. 6 shows that the denoising performance of the proposed method improves with higher wavelet decomposition levels. This effect is most prominent on the PPG signal corrupted by the salt-and-pepper noise. 
This suggests using the maximum level of decomposition (which is 8 in this work). Thus, in the subsequent simulations, we consider an 8-level decomposition. For Fig. 6, mother wavelet used was 'db4'.

Further, the unique morphology of PPG signals prompted us to study the denoising performance of the proposed method for various mother wavelets. Fig. \ref{fig:mse} shows that the MSE of the denoised signal is reduced significantly, compared to the noisy signal for all mother wavelets. However, 'db10' wavelet outperforms all other wavelets for all noise types. 



 \begin{figure} 
\centering
\includegraphics[width=2.8in]{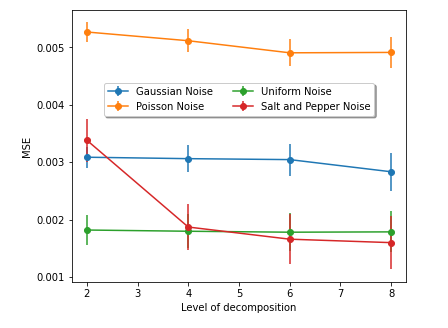}
\caption{Impact of the depth of decomposition on MSE. }
\centering
\label{fig:msedb4}
\end{figure}

\begin{figure} 
\centering
\includegraphics[width=2.8in]{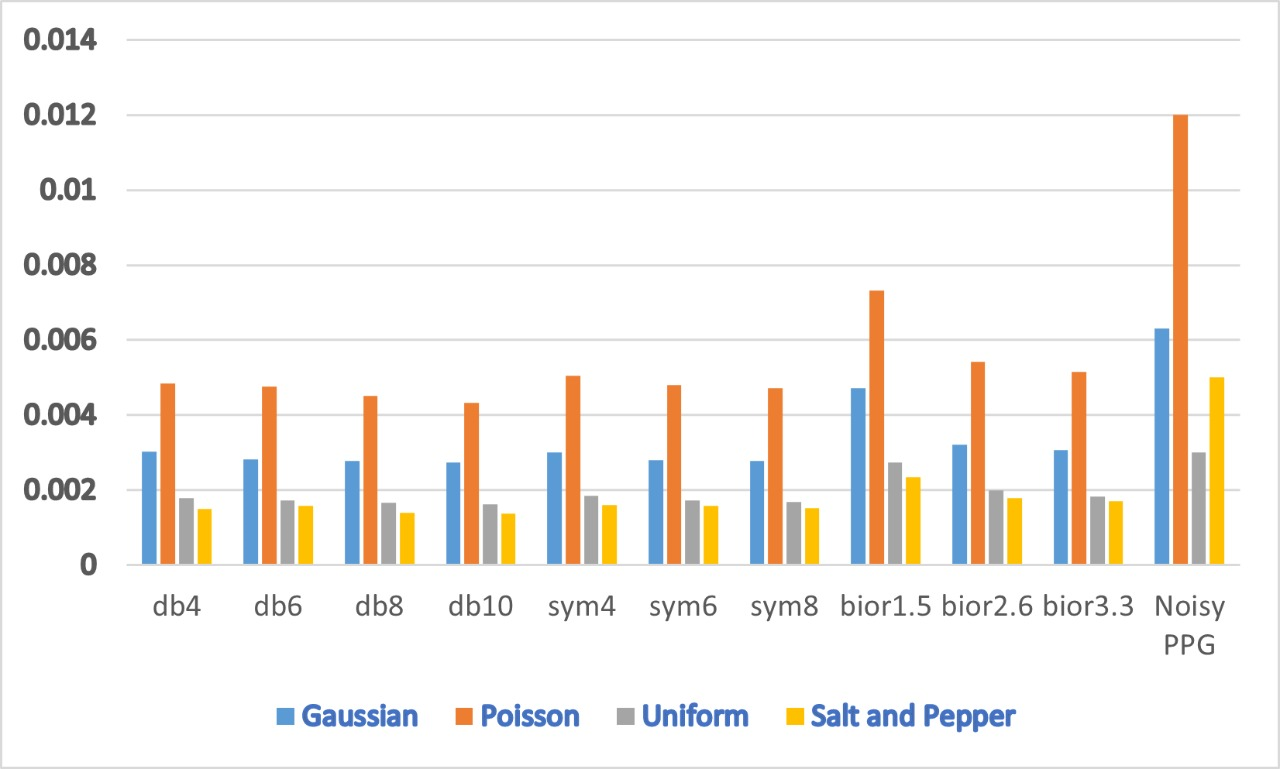}
\caption{MSE of the proposed method under 10 different mother wavelets.}
\centering
\label{fig:mse}
\end{figure}



\begin{figure}
\centering
\includegraphics[width=2.8in]{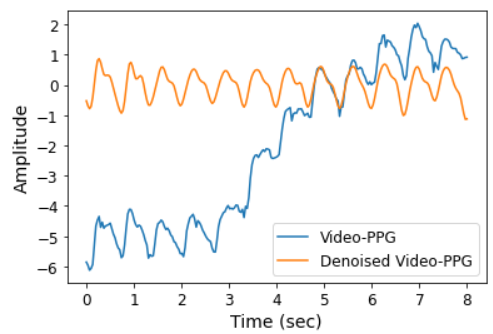}
\caption{Denoised video-PPG signal. }
\centering
\label{fig:NS}
\end{figure}

\textit{Experiment II.}
The second experiment is more ambitious. Here, we utilize clean PPG signals from the BIDMC dataset as before. Additionally, this time we add the artefact signals extracted from our custom video-PPG dataset (by doing FWT decomposition and retaining those detailed coefficients that capture the artefacts) to the training data. This mixed training dataset is then used to train the FFNN of Fig. 4. Note, that for the MSE computation, the clean PPG signal (for the video-PPG data) is obtained by doing the FWT decomposition of the PPG signal, and by discarding those detailed coefficients that captured the artefacts. Once the FFNN is trained, we evaluate its performance by denoise the video-PPG data during the testing phase. Fig. \ref{fig:NS} shows that our proposed method efficiently removes the baseline and high-frequency noise from the noisy PPG signal extracted from the video-PPG data.


\section{Conclusion}
\label{conc}
Under the proposed method, the noisy PPG signal was simultaneously fed to a multi-level FWT decomposition block and a custom FFNN. The FFNN guided the FWT reconstruction phase by systematically setting the wavelet sub-signals corresponding to artefacts and noise to zero.
The FFNN was trained in a supervised manner (using the MSE loss function). Simulation results revealed that our proposed method reduces the MSE of the PPG signal significantly (compared to the MSE of the original noisy PPG signal): by 56.40\% for Gaussian noise, by 64.01\% for Poisson noise, 46.02\% for uniform noise, and by 72.36\% for salt-and-pepper noise (with 'db10' mother wavelet). 



\footnotesize{
\bibliographystyle{IEEEtran}
\bibliography{main}
}
\vfill\break
\end{document}